\documentclass[superscriptaddress,amsmath,amssymb,aps,prl,twocolumn]{revtex4-1}

\usepackage{times}
\usepackage{graphicx}
\usepackage{dcolumn}
\usepackage{bm}
\usepackage{amsmath}
\usepackage{amsfonts}
\usepackage{amssymb}
\usepackage{ulem}
\usepackage[version=3]{mhchem}

\begin{document}

\title{ Subwavelength-thick Lenses with High Numerical Apertures and Large Efficiency Based on High Contrast Transmitarrays}

\author{Amir Arbabi}
\affiliation{T. J. Watson Laboratory of Applied Physics, California Institute of Technology, 1200 E California Blvd., Pasadena, CA 91125, USA}
\author{Yu Horie}
\affiliation{T. J. Watson Laboratory of Applied Physics, California Institute of Technology, 1200 E California Blvd., Pasadena, CA 91125, USA}
\author{Alexander J. Ball}
\affiliation{T. J. Watson Laboratory of Applied Physics, California Institute of Technology, 1200 E California Blvd., Pasadena, CA 91125, USA}
\author{Mahmood Bagheri}
\affiliation{Jet Propulsion Laboratory, California Institute of Technology, Pasadena, CA 91109, USA}
\author{Andrei Faraon}
\email{Corresponding authors: A.F: faraon@caltech.edu, A.A: amir@caltech.edu }
\affiliation{T. J. Watson Laboratory of Applied Physics, California Institute of Technology, 1200 E California Blvd., Pasadena, CA 91125, USA}

\begin{abstract}

We report subwavelength-thick, polarization insensitive micro-lenses  operating  at telecom wavelength with focal spots as small as 0.57 wavelengths  and measured focusing efficiency up to 82\%. The lens design is based on high contrast transmitarrays that enable control of optical phase fronts with subwavelength spatial resolution.  A rigorous method for ultra-thin lens design, and the trade-off between high efficiency and small spot size (or large numerical aperture) are discussed. The transmitarrays, composed of silicon nano-posts on glass, could be fabricated by high-throughput photo or nanoimprint lithography, thus enabling widespread adoption.

\end{abstract}

\maketitle

Flat optical devices thinner than a wavelength promise to replace conventional free-space components for wavefront and polarization control~\cite{Yu2014,Kildishev2013a}. Transmissive flat lenses are particularly interesting for applications in imaging and on-chip optoelectronic integration. Several designs based on metasurfaces~\cite{Yu2011, Aieta2012, Genevet2012,Lin2014a}, high contrast transmitarrays (HCTA) and gratings~\cite{Vo2014},  have been recently implemented but did not provide a performance comparable to conventional curved lenses. Here, for the first time, we report polarization insensitive, micron-thick, HCTA micro-lenses with high numerical apertures and large efficiency.

Flat lenses are most commonly realized with Fresnel structures. However, geometrical complexity of Fresnel lenses and the low efficiency of Fresnel zone plates make them not well suited for integration with wafer-scale processing. Effective medium structures have been proposed as an alternative~\cite{Stork1991,Chen1995, Warren1995,Chen1996,Lalanne1998}, but their deep subwavelength structures and high aspect ratios make their fabrication challenging. Recently, diffractive elements based on plasmonic metasurfaces composed of 2D arrays of ultrathin scatterers have attracted significant attention~\cite{Kildishev2013a,Yu2014,Yu2011,Huang2008,Aieta2012,Ni2013,Aieta2012,Genevet2012,Karimi2014}, but their efficiency is limited to 25\% by fundamental limitations~\cite{Monticone2013} and also suffer from material absorption~\cite{Aieta2012}.

A novel approach is to use high contrast gratings (HCGs), fabricated from semiconductors or high index dielectrics~\cite{Chen2006,Kemiktarak2012,Wu2012a,Lin2014a}, that can be designed with large reflection~\cite{Mateus2004} or transmission~\cite{Lu2010} efficiencies. Wavefront control was originally achieved by rendering one dimensional gratings aperiodic by gradually modifying the local period and duty cycle of the grating~\cite{ Astilean1998, Lu2010, Fattal2010}. Reflecting focusing mirrors were realized using this approach~\cite{Lu2010, Fattal2010,Klemm2013}.

\begin{figure*}[htp]
\centering
\includegraphics[width=2\columnwidth]{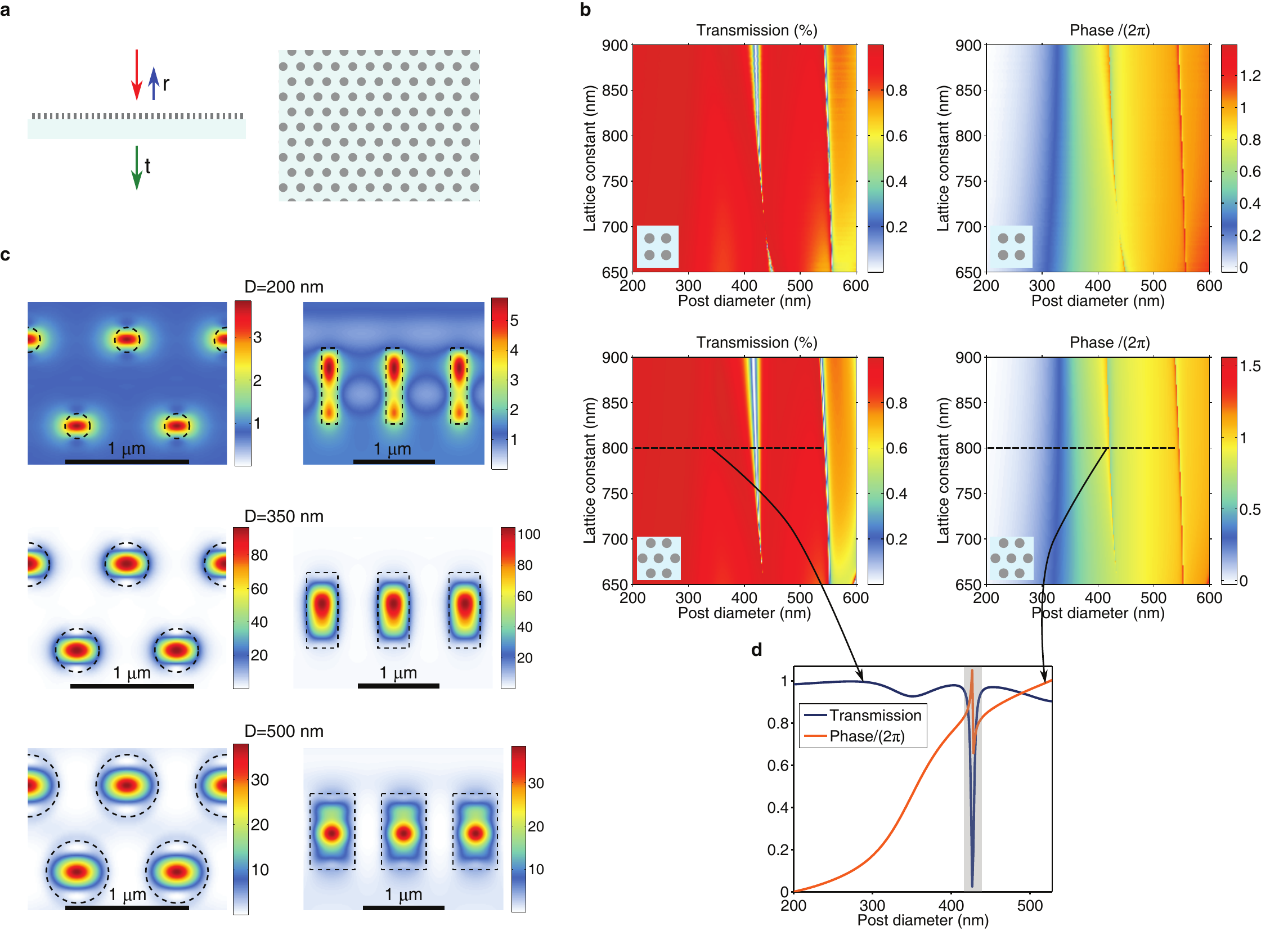}
\caption{\textbf{Simulation results of periodic HCTAs}. \textbf{a} A schematic representation of a periodic HCTA composed of high index posts in a hexagonal lattice (side view on left, top view on right) with transmission coefficient $t$ and reflection coefficient $r$. The posts rest on a low index substrate. \textbf{b}, Simulated transmission and phase of the transmission coefficient for hexagonal and square periodic HCTAs composed of circular amorphous silicon posts on a fused silica substrate as a function of the lattice constant and the post diameter. The insets show the corresponding lattices. \textbf{c}, Top view (on the left) and side view (on the right) of the color coded  magnetic energy density in a periodic HCTA for different post diameters $D$. The dashed black lines depict the boundary of the silicon posts. A plane wave with magnetic energy of 1 is normally incident on the silicon posts from the top. \textbf{d}, Simulated transmission and phase of the transmission coefficient for a family of periodic hexagonal HCTAs with  lattice constant of 800~nm, and varying post diameters.  The shaded part of the graph is excluded when using this graph to map transmission phase to post diameter. In all these simulations, the posts are made of amorphous silicon ($n=3.43$), are 940~nm tall, and the wavelength is $\lambda$=1550~nm.}
\label{fig:lattice_effect}
\end{figure*}

A promising class of aperiodic HCGs can be realized by positioning high index dielectric scatterers on a periodic subwavelength 2D lattice, as shown schematically in Fig.~\ref{fig:lattice_effect}a for a hexagonal lattice of circular posts. Due to subwavelength dimensions, the structure acts as a zero order grating completely described by its transmission and reflection coefficients. The high index results in negligible interaction among scatterers, so the light scattered at each lattice site is dominated by the scatterer proprieties rather than by the collective behavior of multiple coupled scatterers in the lattice. We refer to high contrast zero order gratings composed of disconnected scatterers, which operate in this local scattering regime, as high contrast arrays. The term high contrast transmitarrays (HCTAs) is used when they are designed for large transmission. Low numerical aperture (NA) lenses based on HCTA have been recently reported~\cite{Arbabi2014,Vo2014}.

Here, we explain the HCTA concept and discuss its unique features resulting from the localized scattering phenomena, which enable implementation of diffractive elements with rapidly varying phase profiles. To demonstrate the HCTA versatility, we present design, simulation, fabrication and characterization results of polarization insensitive high NA micro-lenses with high focusing efficacy. Large NA lenses are required in microscopy, high-density data recording, focal plane arrays, and coupling between on-chip photonic components and free-space beams. Current techniques to fabricate on-chip devices that impose rapid phase variation require gray-scale lithography~\cite{Shiono1989,Haruna1990}, a process that is difficult to control. On the other hand, the HCTA devices provide a more reliable alternative with fabrication techniques that lend themselves to wafer-scale processing.

\begin{figure*}[htp]
\centering
\includegraphics[width=2\columnwidth]{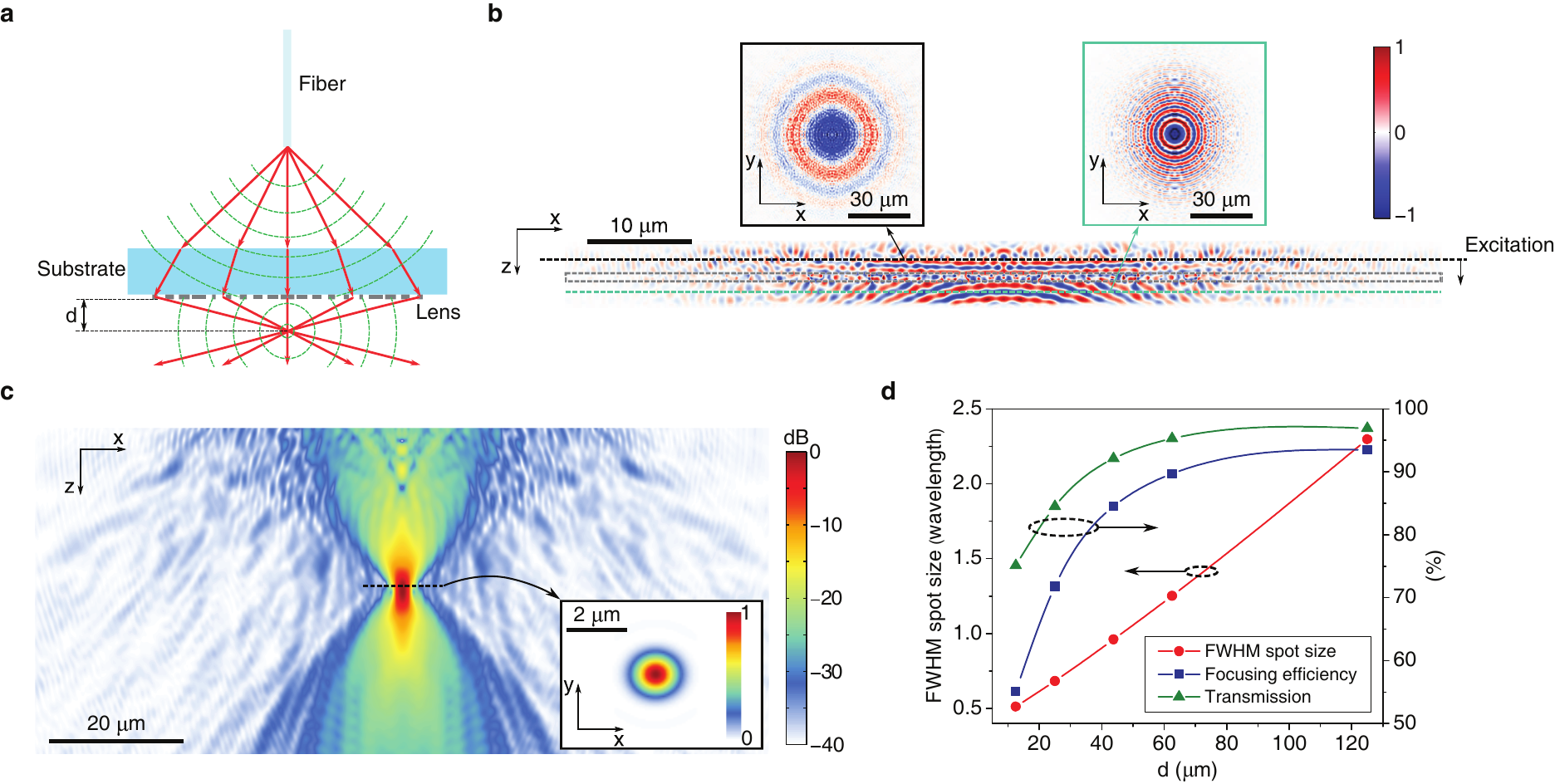}
\caption{\textbf{Simulation results of high NA HCTA micro-lenses}. \textbf{a}, Illustration of high NA focusing of the light from a cleaved optical fiber using an HCTA micro-lens. \textbf{b}, Electric field distribution at the \textit{xz} cross section, in the excitation plane (inset), and immediately after passing through the micro-lens (inset). \textbf{c}, Logarithmic scale electric energy density in the \textit{xz} cross section. the inset shows the real part of the z component of the Poynting vector at the plane of focus. \textbf{d}, Simulated plane of focus FWHM spot size, transmission, and focusing efficiency of the high NA HCTA micro-lenses for devices with varying focusing distances. The simulated points are shown by the symbols and the solid lines are eye guides. All the devices simulated in this figure are a factor of four smaller than the devices fabricated and measured in Fig.~\ref{fig:fabrication} and Fig.~\ref{fig:Hi_NA_results}}.
\label{fig:high_NA_simulation}
\end{figure*}

Figure~\ref{fig:lattice_effect}b shows the simulated transmission and phase of the transmission coefficient for two gratings composed of circular amorphous silicon posts arranged in hexagonal and rectangular lattices, as functions of lattice constant and post diameter (See Fig.~\ref{fig:lattice_effect} legend for dimensions). The gratings are designed to operate at $\lambda$=1550~nm, but the concept is scalable to any wavelength.  In each case, the amplitude and phase depend primarily on the post diameter, which is an indication that, in this parameter regime, the scattering is a local effect and is not affected significantly by coupling among scatterers. This is also confirmed by the almost identical transmission properties of the hexagonal and square lattice. Examining the near field distribution of the grating reveals the underlying physical mechanism behind the local scattering effect. As shown in Fig.~\ref{fig:lattice_effect}c, light is concentrated inside the posts that behave as weakly coupled low quality factor resonators. This behavior is fundamentally different from the low contrast gratings operating in the effective medium regime whose diffractive characteristics are mainly determined by the duty cycle and the filling factor. We also note that the structures created by changing the bar width and period of 1D HCGs~\cite{Fattal2010, Lu2010, Klemm2013} are not considered HCTA since there is a strong coupling along the bar direction and a rapid variation of the local transmission or reflection properties of the  structure is not achievable along that direction.

\begin{figure*}[htp]
\centering
\includegraphics[width=2\columnwidth]{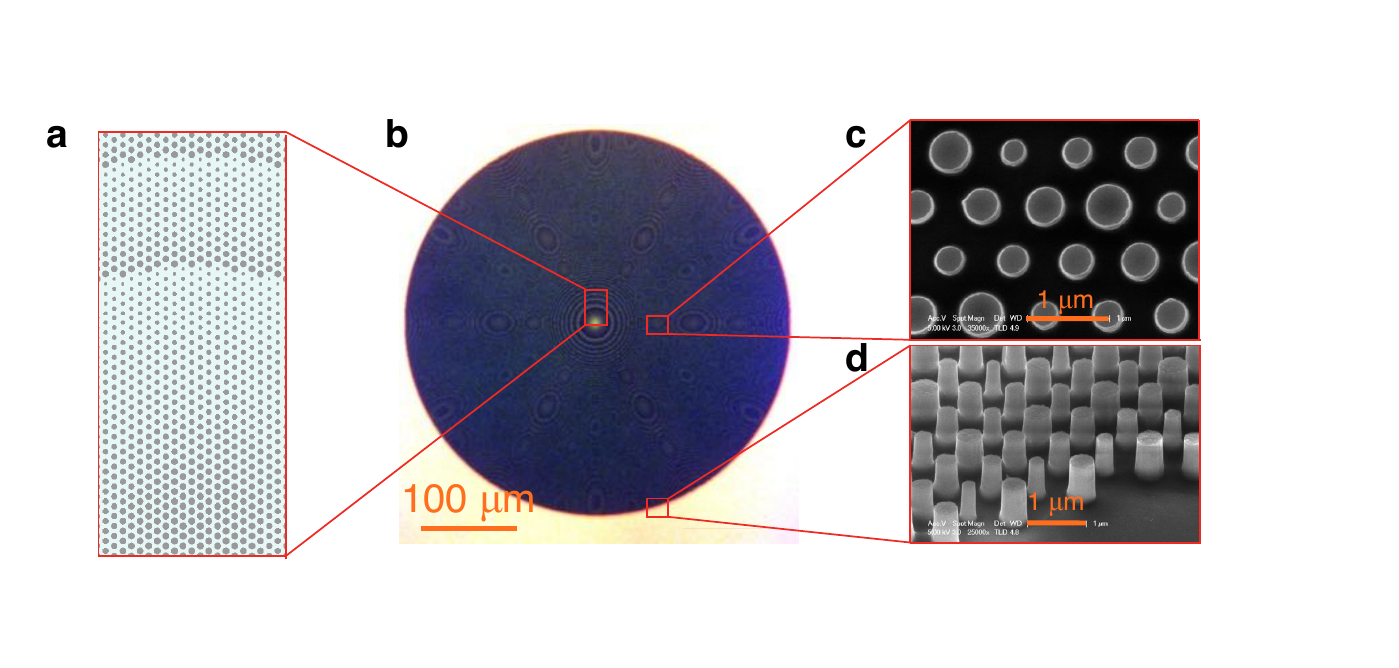}
\caption{\textbf{Schematic illustration and images of fabricated HCTA lenses}. \textbf{a}, Schematic of the aperiodic HCTA used in the high NA micro-lens. \textbf{b}, Optical microscope image of a fabricated HCTA lens with large NA.  \textbf{c,d}, Scanning electron microscope images of the silicon posts forming the HCTA micro-lens.}
\label{fig:fabrication}
\end{figure*}

To design an HCTA that implements a transmissive phase mask, we find a family of periodic HCTAs with the same lattice but with different scatterers that provide large transmission amplitude while the phase spans the entire 0 to $2\pi$ range. Such a family is shown in Figure~\ref{fig:lattice_effect}d where the post diameter is varied from 200~nm to 550~nm in a hexagonal lattice with 800~nm period while transmission is greater than 92\%. To create the phase mask we start from an empty lattice and at each lattice site we place a scatterer from the periodic HCTAs that most closely imparts the desired phase change onto the transmitted light.  Any arbitrary transmissive phase masks can be realized using this method. An example phase mask for a transmissive micro-lens is shown in Fig.~\ref{fig:fabrication}a.  To minimize scattering from aperiodic HCTAs into non-zero orders, a gradual change in the scatterer size is preferable.

The unprecedented possibility to realize any transmissive masks using HCTAs enables the implementation of micro-lenses with exotic phase profiles, optimized for specific tasks, such as large NA lenses. To design these components, the conventional ray tracing technique is not applicable. A general rigorous technique for determining the optimum transmissive mask to shape an incident optical wavefront to a desired form is given in Supplementary Information, S.1. Using this technique and the HCTAs in Fig.~\ref{fig:lattice_effect}d, we found the optimum phase mask for micro-lenses that focus $\lambda$=1550~nm light from a single mode fiber to the smallest spot.  We designed a set of 400~$\mu$m diameter high NA micro-lenses that focus the light from a single mode fiber located 600~$\mu$m away from the backside of the substrate (500~$\mu$m thick fused silica) to points located at distances ranging from $d$=50~$\mu$m to $d$=500~$\mu$m away from the micro-lenses (Fig.~\ref{fig:high_NA_simulation}a). We refer to $d$ as the focusing distance.

The lens performance was evaluated by full 3D finite difference time domain (FDTD) simulations~\cite{Oskooi2010b}. To reduce the simulation size, micro-lenses with the same NA but with a factor of four smaller dimensions (100~$\mu$m diameter, 150~$\mu$m spacing between the illuminating fiber and the back side of the 125~$\mu$m thick substrate)   were simulated. Figures~\ref{fig:high_NA_simulation}b and c show the results for a micro-lens that focuses at $d$=25~$\mu$m away from the lens. The full width at half maximum (FWHM) of the focal spot is 1.06~$\mu$m or 0.68$\lambda$ ($\lambda$=1550~nm). The simulation indicates that 85\% of the incident light is transmitted by the lens and 72\% is concentrated in the focus (See Methods for details).

Simulated values of the transmission, focusing efficiency and FWHM spot size for several micro-lenses with $d$ ranging from 12.5~$\mu$m to 125~$\mu$m are presented in Fig.~\ref{fig:high_NA_simulation}d. Higher NAs and smaller spot sizes correlate with decreased transmission and lower focusing efficiencies. This is due to the rapidly varying phase profile close to the circumference of high NA lenses, which is under-sampled by the HCTA lattice (shrinking the lattice constant of the HCTA reduces this trade-off, see Supplementary Information S.2). The micro-lens with $d$=12.5~$\mu$m has a simulated FWHM spot size of 0.51$\lambda$ which is close to the smallest possible diffraction limited value of 0.5$\lambda$. The relatively high efficiency (more than 50\%) and the diffraction limited focusing of this micro-lens confirms the validity of our technique for determining the optimum phase profile, and demonstrates an example of the high performance that can be achieved by HCTA flat diffractive elements.

\begin{figure*}[htp]
\centering
\includegraphics[width=2\columnwidth]{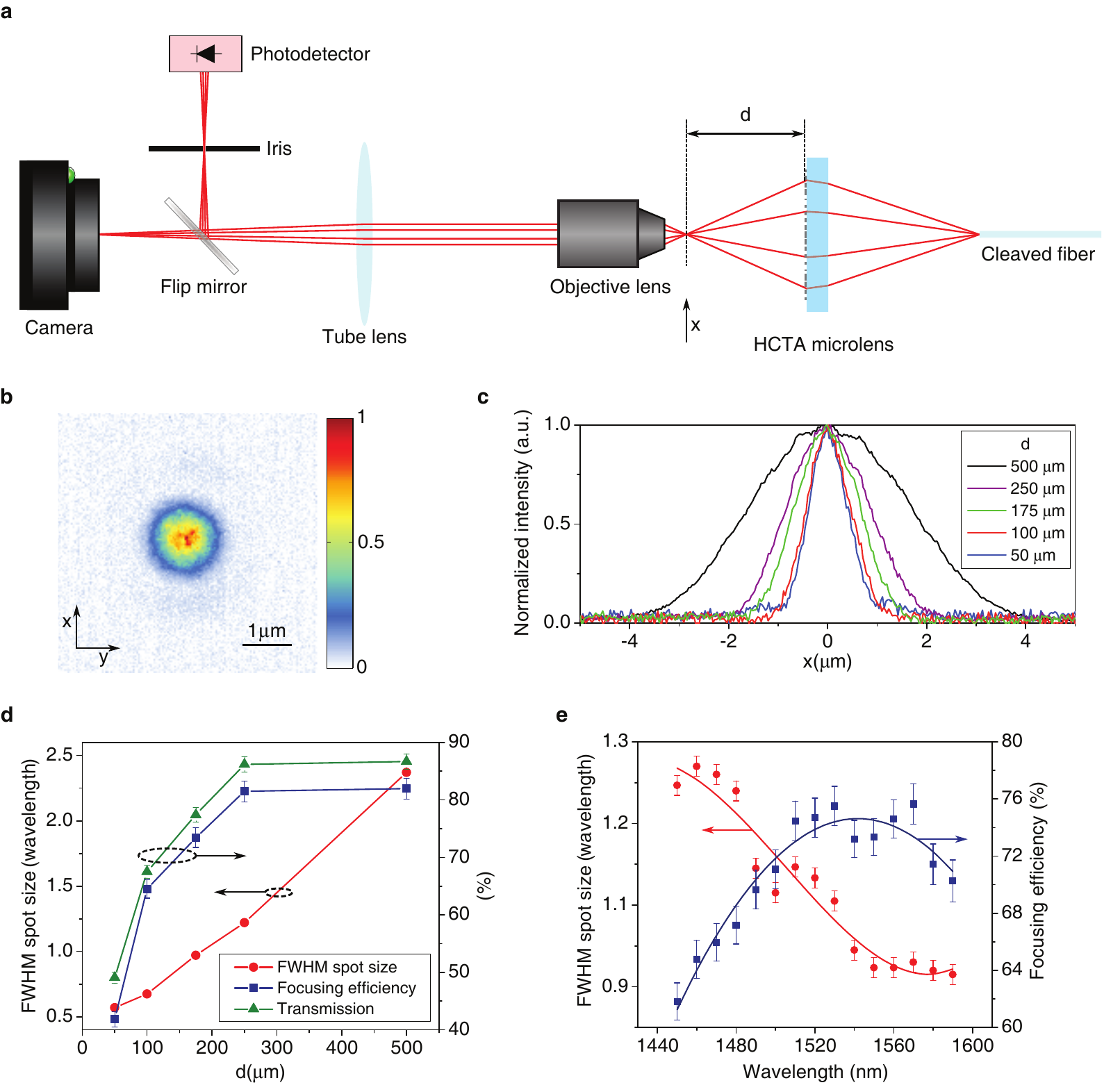}
\caption{\textbf{Measurement results of high NA HCTA micro-lenses}. \textbf{a}, Measurement setup for imaging the focal spot size and measuring the efficiency of the high NA HCTA micro-lenses. The flip mirror was inserted into the setup only during efficiency measurements. \textbf{b}, Measured 2D intensity profile at the plane of focus for a micro-lens with $d$=50~$\mu$m. \textbf{c}, Normalized measured intensity profiles of high NA micro-lenses with different focal lengths at their plane of focus. \textbf{d}, Measured plane of focus FWHM spot size, transmission, and focusing efficiency of the HCTA micro-lenses as a function of their focusing distance. The measurement data are represented by the symbols and the solid lines are eye guides. \textbf{e}, Wavelength dependence of the FWHM spot size, transmission, and focusing efficiency for the micro-lens with $d$=175~$\mu$m.}
\label{fig:Hi_NA_results}
\end{figure*}

The high NA lenses were fabricated in a hydrogenated amorphous silicon film deposited on a 500~$\mu$m thick fused silica substrate as described in the Methods section. Images of the fabricated devices are shown in Fig.~\ref{fig:fabrication}a-c. The characterization was performed in a setup (Fig.~\ref{fig:Hi_NA_results}a) consisting of a custom built microscope that images the plane of focus of the micro-lens (See Methods). The micro-lenses were illuminated with 1550~nm light emitted from a cleaved single mode fiber positioned 600~$\mu$m away from the substrate backside. The normalized measured intensity profile at the plane of focus for a micro-lens with the focusing distance of $d$=50~$\mu$m is shown in Fig.~\ref{fig:Hi_NA_results}b. The intensity profiles for micro-lenses with different focusing distance are plotted in Fig.~\ref{fig:Hi_NA_results}c.

Figure~\ref{fig:Hi_NA_results}d shows the measured FWHM spot size, transmission, and focusing efficiency for devices with different focusing distances. The micro-lens designed for $d$=50~$\mu$m focuses light to a 0.57$\lambda$ FWHM spot size, and the micro-lens designed for $d$=500~$\mu$m shows more than 82\% focusing efficiency. These results agree well with the simulation results presented in Fig.~\ref{fig:high_NA_simulation}d, although the measured focusing efficiencies are ~10\% smaller (3\% is attributed to reflection from substrate backside interface and 7\% to scattering by the random roughness of the etched silicon posts). As expected from simulations (Fig.~\ref{fig:high_NA_simulation}d), the measured  focusing efficiency decreases as the NA increases.    

The wavelength dependence of the FWHM spot size and focusing efficiency of a micro-lens with $d$=175~$\mu$m are presented in Fig.~\ref{fig:Hi_NA_results}e. The FWHM spot size increases slightly at shorter wavelength, and the focusing efficiency reduces by $\sim$5~\% at 50~nm away from the design wavelength. Also, by changing the laser wavelength from 1550~nm to 1450~nm, the focusing distance changed from $\sim$175~$\mu$m to $\sim$195~$\mu$m.

In conclusion, the HCTAs enable shaping of the wavefront of light at will, efficiently, and with sub-wavelength resolution. The exceptional freedom provided in the implementation of any desired masks allows for achieving the best performance for any particular functionality. The HCTA micro-lenses with focusing efficiencies up to 82\%, and FWHM spot sizes down to 0.57$\lambda$, to the best of our knowledge, represent the best performance among any types of flat high NA micro-lens experimentally reported so far.
Combined with their planar form factor, these structures will enable on-chip optical systems created by cascading multiple diffractive elements. One recent demonstration is a planar retroreflector integrating a HCTA lens and a reflectarray focusing mirror   ~\cite{Arbabi2014a}. We envision near future application of HCTA based devices in realization of more complex optical systems with new functionalities.

\clearpage

\section*{Methods}

\textbf{Sample Fabrication}. The HCTA pattern was defined in ZEP520A positive resist using a Vistec EBPG5000+ electron beam lithography system. After developing the resist, the pattern was transferred into a 70 nm thick aluminum oxide layer deposited by electron beam evaporation using the lift-off technique. The patterned aluminum oxide served as hard mask for the subsequent dry etching of the 940~nm thick silicon layer in a mixture of \ce{C_4F_8} and \ce{SF_6} plasma.

\textbf{Measurement procedure}. The microscope uses a 100X objective (Olympus UMPlanFl) with the NA of 0.95, and a tube lens (Thorlabs LB1945-C) with focal distance of 20~cm which is anti-reflection coated for the 1050-1620~nm wavelength range. The magnification of the microscope was found by imaging a calibration sample with known feature dimensions.

The transmission and focusing efficiency of the micro-lenses were measured by inserting a flip mirror (as shown in Fig.~\ref{fig:Hi_NA_results}a) in front of the camera. To measure the optical power focused by the micro-lens, the active area of the photodetector (Newport 818-IR) was reduced using an iris. The radius of the iris aperture was adjusted to three times of the measured FWHM spot size of the micro-lens on the camera. The total transmitted power was measured by opening the iris aperture completely. The total power incident on the microlens was measured by removing the micro-lens from the setup and bringing the fiber tip into the focus of the microscope. The non-uniformities seen in the intensity profile in Fig.~\ref{fig:Hi_NA_results} are due to the local variations in the sensitivity of the camera (Digital CamIR 1550 by Applied Scintillation Technologies) and are observed even when directly imaging the light from an optical fiber.

\textbf{Simulations}. We found the electric and magnetic fields of the light from the fiber on a plane close to the lens using the plane wave expansion technique. Then, these fields were used to determine the equivalent electric and magnetic surface current densities, which were used as excitation sources in the FDTD simulation. This allowed us to reduce the size of the simulation domain to a smaller volume surrounding the micro-lens.

In Fig.~\ref{fig:high_NA_simulation} , the FWHM spot size is found by fitting a 2D Gaussian function to the z component of the Poynting vector at the plane of focus (shown in Fig.~\ref{fig:high_NA_simulation}c). The focusing efficiency is defined as the fraction of the incident light that passes through a circular aperture in the plane of focus with a radius equal to three times the FWHM spot size

\section*{Acknowledgement}
This work was supported by Caltech/JPL president and director fund (PDF) . Amir Arbabi was also supported by DARPA. Yu Horie was supported by JASSO fellowship and the  ÒLight-Material Interactions in Energy ConversionÓ Energy Frontier Research Center funded
by the US Department of Energy, Office of Science, Office of Basic Energy Sciences under Award No. DE-SC0001293. Alexander Ball was supported by the Summer Undergraduate Research Fellowship (SURF) at Caltech. The device nanofabrication was performed in the Kavli Nanoscience Institute at Caltech.

The authors thank to David Fattal and Sonny Vo for useful discussion.

\clearpage

\newcommand{\beginsupplement}{%
        \setcounter{table}{0}
        \renewcommand{\thetable}{S\arabic{table}}%
        \setcounter{figure}{0}
        \renewcommand{\thefigure}{S\arabic{figure}}%
     }

\onecolumngrid

      \beginsupplement

\section{Supplementary Information for ``Subwavelength-thick Lenses with High Numerical Apertures and Large Efficiency Based on High Contrast Transmitarrays"}

\subsection{S.1 Optimum Transmissive Mask Design}

\begin{figure*}[htp]
\centering
\includegraphics[width=1\columnwidth]{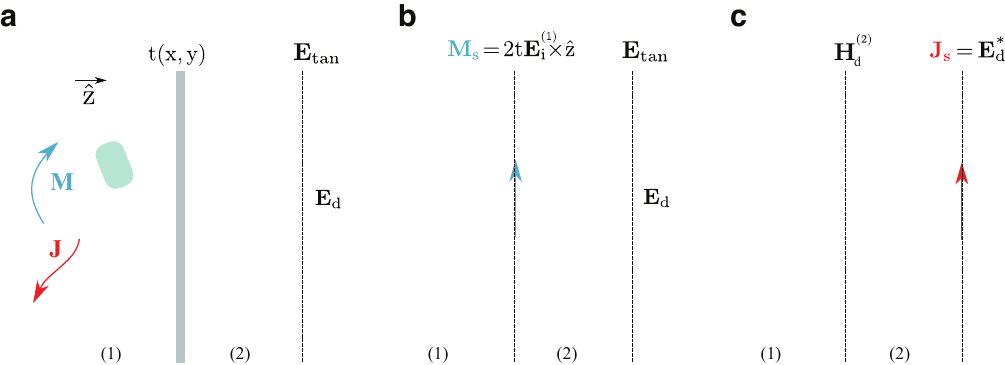}
\caption{\textbf{Optimum transmissive mask design for shaping an incident light to the desired tangential form}. \textbf{a}, The light from the sources and scatterers in half space (1) passes through the transmission mask $t(x,y)$, and its tangential electric field on the target plane is represented by $\mathbf{E}_\mathrm{tan}$. \textbf{b}, The equivalent magnetic surface current density $\mathbf{M}_\mathrm{s}$ emits the same tangential electric field $\mathbf{E}_\mathrm{tan}$ on the target plane. \textbf{c}, The magnetic field $\mathbf{H}_\mathrm{d}^{(2)}$ is emitted by an electric surface current density $\mathbf{J}_\mathrm{s}$  which is located on the target plane.}
\label{fig:optimum_mask}
\end{figure*}

We consider the general case of using a transmissive mask to modify the optical wavefront emitted by given sources to a desired form. The light is generated by sources located in half space (1) as shown schematically in Fig.~\ref{fig:optimum_mask}a. The tangential component of the electric field of the incident light just before propagating through the transmissive mask is represented by $\mathbf{E}_\mathrm{i}^{(1)}$. The incident field might be, for example, a diverging beam from a semiconductor laser or a collimated Gaussian beam. The desired output wavefront can be chosen arbitrarily (examples include a beam that is matched to a mode of an optical fiber, a Bessel beam, or a tightly focused beam). Since the propagation is governed by Maxwell's equations, the desired output beam is fully described by the tangential components of its electric field on a target plane parallel to the transmissive mask. We represent this desired tangential component by $\mathbf{E}_\mathrm{d}$ (as shown in Fig.~\ref{fig:optimum_mask}a)).

The output beam formed by the transmissive mask is in general different from the desired beam, and its tangential electric field on the target plane $\mathbf{E}_\mathrm{tan}$ is an approximation for the desired tangential electric field. Our main objective is to determine the transmissive mask $t(x,y)$ such that $\mathbf{E}_\mathrm{tan}$ is the best possible approximation for the $\mathbf{E}_\mathrm{d}$. A useful measure for quantifying the accuracy of the approximation is the norm of the projection integral defined as 
\begin{equation}
\left|<\mathbf{E}_\mathrm{tan},\mathbf{E}_\mathrm{d}>\right|=\left|\int \mathbf{E}_\mathrm{tan}\cdot\mathbf{E}^*_\mathrm{d}\mathrm{d}s\right|,
\label{eq:similarity_integral}
\end{equation}
where $*$ represents the complex conjugate operation; and the surface integral is evaluated over the target plane. For the best approximation, this norm should be maximized.

To relate the projection integral to the transmission mask we use the equivalence principle and the reciprocity theorem. According to the definition of a transmissive mask, the tangential electric field of the light just after it passes through the transmissive mask is given by $t\mathbf{E}_\mathrm{i}$. Using the equivalence principle, a magnetic surface current density $\mathbf{M}_\mathrm{s}=2t\mathbf{E}_\mathrm{i}\times\hat{z}$ located at the output plane of the transmissive mask and emitting in vacuum, will generate the same beam in the region (2) as the original sources and the transmissive mask (see Fig~\ref{fig:optimum_mask}b). Now we consider an electric surface current density with $\mathbf{J}_s=\mathbf{E}^*_\mathrm{d}$ on the target plane which is emitting in vacuum. We refer to the magnetic field emited by $\mathbf{J}_\mathrm{s}$ at the output plane of the transmissive mask by $\mathbf{H}_\mathrm{d}^{(2)}$. Using the reciprocity theorem~\cite{Harrington2001}, we can write  
\begin{equation}
\int\mathbf{E}_\mathrm{tan}\cdot\mathbf{J}_\mathrm{s}\mathrm{d}s=\int\mathbf{H}_\mathrm{d}^{(2)}\cdot\mathbf{M}_\mathrm{s}\mathrm{d}s.
\label{eq:reciprocity}
\end{equation}
From~(\ref{eq:similarity_integral}) and~(\ref{eq:reciprocity}) we obtain 
\begin{equation}
\left|<\mathbf{E}_\mathrm{tan},\mathbf{E}_\mathrm{d}>\right|=2\left|\int t\mathbf{H}_\mathrm{d}^{(2)}\cdot(\mathbf{E}_\mathrm{i}^{(1)}\times\hat{z})\mathrm{d}s\right|.
\label{eq:optimum_mask_eq2}
\end{equation}
Using~(\ref{eq:optimum_mask_eq2}), we see that the best approximation to a desired output beam is achieved when $|t|=1$ and
\begin{equation}
\angle t=-\angle\left( \mathbf{H}_\mathrm{d}^{(2)}\cdot(\mathbf{E}_\mathrm{i}^{(1)}\times\hat{z})\right).
\label{eq:optimum_phase}
\end{equation}
In other words, the best transmissive mask is a phase mask, and we can determine its phase profile as follows. First, we find the tangential component of the incident light at the location of the transmissive mask ($\mathbf{E}_\mathrm{i}^{(1)}$); this can be done either analytically or numerically depending on the type of the excitation. Next, we consider an electric surface current density with the same spatial distribution as the desired tangential electric field at the target plane which is emitting in the free space. We find the magnetic field emitted by this current at the location of the transmission mask ($\mathbf{H}_\mathrm{d}^{(2)}$). Since $\mathbf{J}_\mathrm{s}$ is planar and is radiating in free space, its fields can be obtained using a simple method such as the plane wave expansion technique~\cite{Born1999}. Finally, we obtain the phase profile of the transmission mask using~(\ref{eq:optimum_phase}).
To achieve a lens with tight focus, the desired field should be set to a uniform electric field confined inside a circle with deep subwavelength radius located at the plane of focus.

\subsection{S.2 Under-sampling of the phase profile}

Considering the micro-lens in the geometrical optics picture offers an intuitive understanding of how the under-sampling of the phase profile affects the lens performance. According to the geometrical optics, a ray that is propagating parallel to the lens's optical axis is deflected by the lens toward its focal point. The rays that are propagating farther away from the optical axis are deflected by larger angles, and the lens's NA represents the sine of the largest deflection angle. Therefore, a lens can be considered as a deflector whose local deflection angle gradually increases from zero at the center of the lens to its maximum (which is given by $\sin^{{-1}}(\mathrm{NA})$) at the perimeter of the lens. We show that due to the under sampling of the phase profile by the HCTA lattice, the deflection efficiency of the HCTA deflectors decreases by increasing their deflection angle. The lower deflection efficiency at larger deflection angles leads to the lower focusing efficiency of the micro-lenses with higher NA. 

\begin{figure*}[htp]
\centering
\includegraphics[width=1\columnwidth]{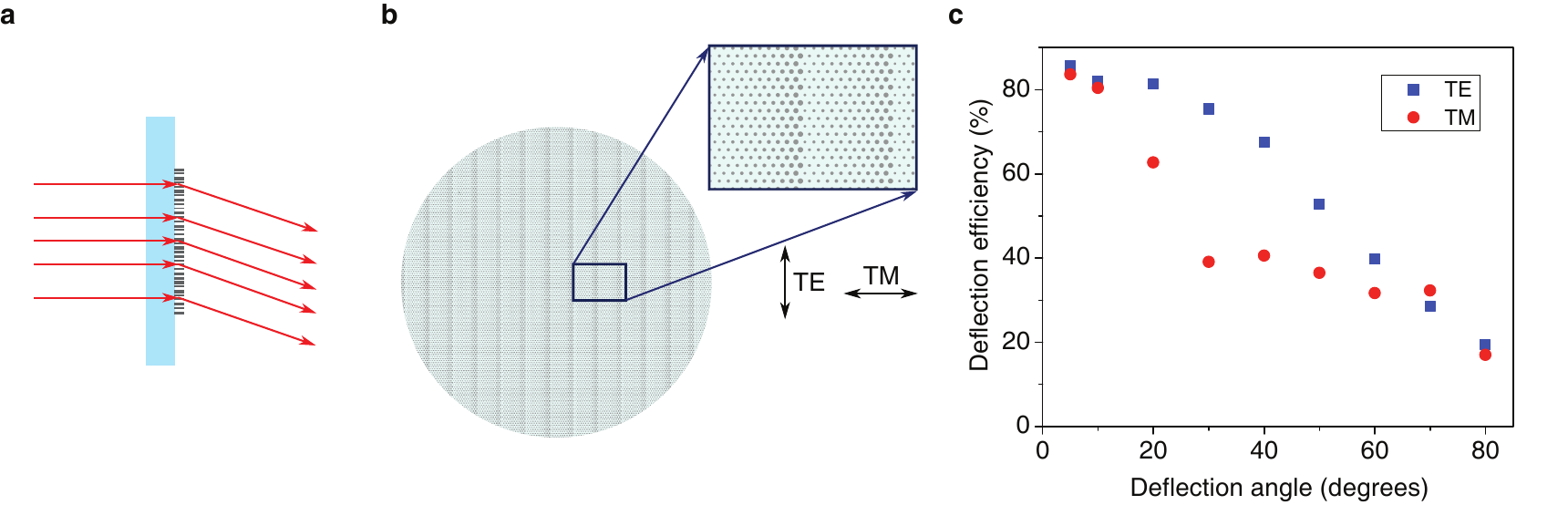}
\caption{\textbf{Measurement results of HCTA beam deflectors.} \textbf{a}, Schematic illustration of an ideal beam deflector. \textbf{b}, The HCTA pattern that implements a uniform phase ramp along the horizontal direction. The polarization direction for the TE and TM polarizations are also shown. \textbf{c}, Measured deflection efficiency of a set of beam deflectors for the transverse electric and magnetic polarized incident beam as a function of deflection angle.}
\label{fig:deflector_efficiency}
\end{figure*}

A uniform deflector functions similar to a blazed grating and deflects a monochromatic normally incident light by a fixed angle (as shown in Fig.~\ref{fig:deflector_efficiency}a). The uniform deflector has a linearly varying phase profile whose slope is proportional to the sine of its deflection angle. A schematic illustration of an HCTA uniform deflector, with phase profile varying linearly along the horizontal direction, is depicted in Fig.~\ref{fig:deflector_efficiency}b. Nine 400~$\mu$m diameter uniform deflectors with different deflection angles were fabricated using the same family of periodic HCTAs used for the high NA micro-lenses and the same fabrication process.

The deflectors were illuminated with a linearly polarized collimated laser beam with beam radius of approximately 100~$\mu$m. The deflected power was measured using a photodetector located 10~cm away from the deflector along the expected deflection direction. The deflection efficiency was obtained by dividing the measured deflected power by the incident power. The measured deflection efficiencies for two linear orthogonal polarizations are depicted in Fig.~\ref{fig:deflector_efficiency}c. The directions of polarization for the TE and TM polarized lights are shown in Fig.~\ref{fig:deflector_efficiency}b. The TE polarization corresponds to the transverse electric polarized deflected light, while the deflected light is transverse magnetic polarized for the TM polarization. 

As Fig.~\ref{fig:deflector_efficiency}c shows, the deflection efficiency of an HCTA deflector decreases as its deflection angle increases. The deflection efficiency drop is faster for the TM polarized incident light compared to the TE one. We attribute the efficiency reduction to the under sampling of the phase profile of the deflectors with large deflection angles. The desired phase profile of a deflector with deflection angle of $\theta$ is sampled by $n=\lambda/(a\sin(\theta))$ unit cells over $2\pi$ phase variation, where $a$ is the lattice constant of the HCTA. For the HCTA used in this study, the lattice constant is roughly equal to a half of a wavelength; therefore, for a 40$^\circ$ deflector, the full phase range of $2\pi$ is sampled by approximately three unit cells. Similar diffraction efficiency reduction due to phase sampling and quantization error is encountered in the design and implementation of Fresnel lenses with a limited number of levels~\cite{Wyrowski1992}. One approach to increase the efficiency of a high NA micro-lens is to use an HCTA with a smaller lattice constant. This leads to a finer sampling of the desired micro-lens profile. For example, as Fig.1b,d in the main manuscript show, we could use the lattice constant of 650~nm instead of 800~nm and still achieve the full $2\pi$ transmission phase range by changing the post diameters.


\begin{thebibliography}{29}%
\makeatletter
\providecommand \@ifxundefined [1]{%
 \@ifx{#1\undefined}
}%
\providecommand \@ifnum [1]{%
 \ifnum #1\expandafter \@firstoftwo
 \else \expandafter \@secondoftwo
 \fi
}%
\providecommand \@ifx [1]{%
 \ifx #1\expandafter \@firstoftwo
 \else \expandafter \@secondoftwo
 \fi
}%
\providecommand \natexlab [1]{#1}%
\providecommand \enquote  [1]{``#1''}%
\providecommand \bibnamefont  [1]{#1}%
\providecommand \bibfnamefont [1]{#1}%
\providecommand \citenamefont [1]{#1}%
\providecommand \href@noop [0]{\@secondoftwo}%
\providecommand \href [0]{\begingroup \@sanitize@url \@href}%
\providecommand \@href[1]{\@@startlink{#1}\@@href}%
\providecommand \@@href[1]{\endgroup#1\@@endlink}%
\providecommand \@sanitize@url [0]{\catcode `\\12\catcode `\$12\catcode
  `\&12\catcode `\#12\catcode `\^12\catcode `\_12\catcode `\%12\relax}%
\providecommand \@@startlink[1]{}%
\providecommand \@@endlink[0]{}%
\providecommand \url  [0]{\begingroup\@sanitize@url \@url }%
\providecommand \@url [1]{\endgroup\@href {#1}{\urlprefix }}%
\providecommand \urlprefix  [0]{URL }%
\providecommand \Eprint [0]{\href }%
\providecommand \doibase [0]{http://dx.doi.org/}%
\providecommand \selectlanguage [0]{\@gobble}%
\providecommand \bibinfo  [0]{\@secondoftwo}%
\providecommand \bibfield  [0]{\@secondoftwo}%
\providecommand \translation [1]{[#1]}%
\providecommand \BibitemOpen [0]{}%
\providecommand \bibitemStop [0]{}%
\providecommand \bibitemNoStop [0]{.\EOS\space}%
\providecommand \EOS [0]{\spacefactor3000\relax}%
\providecommand \BibitemShut  [1]{\csname bibitem#1\endcsname}%
\let\auto@bib@innerbib\@empty
\bibitem [{\citenamefont {Yu}\ and\ \citenamefont {Capasso}(2014)}]{Yu2014}%
  \BibitemOpen
  \bibfield  {author} {\bibinfo {author} {\bibfnamefont {N.}~\bibnamefont
  {Yu}}\ and\ \bibinfo {author} {\bibfnamefont {F.}~\bibnamefont {Capasso}},\
  }\href {\doibase 10.1038/NMAT3839} {\bibfield  {journal} {\bibinfo  {journal}
  {Nature materials}\ }\textbf {\bibinfo {volume} {13}},\ \bibinfo {pages}
  {139} (\bibinfo {year} {2014})}\BibitemShut {NoStop}%
\bibitem [{\citenamefont {Kildishev}\ \emph {et~al.}(2013)\citenamefont
  {Kildishev}, \citenamefont {Boltasseva},\ and\ \citenamefont
  {Shalaev}}]{Kildishev2013a}%
  \BibitemOpen
  \bibfield  {author} {\bibinfo {author} {\bibfnamefont {A.~V.}\ \bibnamefont
  {Kildishev}}, \bibinfo {author} {\bibfnamefont {A.}~\bibnamefont
  {Boltasseva}}, \ and\ \bibinfo {author} {\bibfnamefont {V.~M.}\ \bibnamefont
  {Shalaev}},\ }\href {\doibase 10.1126/science.1232009} {\bibfield  {journal}
  {\bibinfo  {journal} {Science (New York, N.Y.)}\ }\textbf {\bibinfo {volume}
  {339}},\ \bibinfo {pages} {1232009} (\bibinfo {year} {2013})}\BibitemShut
  {NoStop}%
\bibitem [{\citenamefont {Yu}\ \emph {et~al.}(2011)\citenamefont {Yu},
  \citenamefont {Genevet}, \citenamefont {Kats}, \citenamefont {Aieta},
  \citenamefont {Tetienne}, \citenamefont {Capasso},\ and\ \citenamefont
  {Gaburro}}]{Yu2011}%
  \BibitemOpen
  \bibfield  {author} {\bibinfo {author} {\bibfnamefont {N.}~\bibnamefont
  {Yu}}, \bibinfo {author} {\bibfnamefont {P.}~\bibnamefont {Genevet}},
  \bibinfo {author} {\bibfnamefont {M.~a.}\ \bibnamefont {Kats}}, \bibinfo
  {author} {\bibfnamefont {F.}~\bibnamefont {Aieta}}, \bibinfo {author}
  {\bibfnamefont {J.-P.}\ \bibnamefont {Tetienne}}, \bibinfo {author}
  {\bibfnamefont {F.}~\bibnamefont {Capasso}}, \ and\ \bibinfo {author}
  {\bibfnamefont {Z.}~\bibnamefont {Gaburro}},\ }\href {\doibase
  10.1126/science.1210713} {\bibfield  {journal} {\bibinfo  {journal} {Science
  (New York, N.Y.)}\ }\textbf {\bibinfo {volume} {334}},\ \bibinfo {pages}
  {333} (\bibinfo {year} {2011})}\BibitemShut {NoStop}%
\bibitem [{\citenamefont {Aieta}\ \emph {et~al.}(2012)\citenamefont {Aieta},
  \citenamefont {Genevet}, \citenamefont {Kats}, \citenamefont {Yu},
  \citenamefont {Blanchard}, \citenamefont {Gaburro},\ and\ \citenamefont
  {Capasso}}]{Aieta2012}%
  \BibitemOpen
  \bibfield  {author} {\bibinfo {author} {\bibfnamefont {F.}~\bibnamefont
  {Aieta}}, \bibinfo {author} {\bibfnamefont {P.}~\bibnamefont {Genevet}},
  \bibinfo {author} {\bibfnamefont {M.~A.}\ \bibnamefont {Kats}}, \bibinfo
  {author} {\bibfnamefont {N.}~\bibnamefont {Yu}}, \bibinfo {author}
  {\bibfnamefont {R.}~\bibnamefont {Blanchard}}, \bibinfo {author}
  {\bibfnamefont {Z.}~\bibnamefont {Gaburro}}, \ and\ \bibinfo {author}
  {\bibfnamefont {F.}~\bibnamefont {Capasso}},\ }\href {\doibase
  10.1021/nl302516v} {\bibfield  {journal} {\bibinfo  {journal} {Nano letters}\
  }\textbf {\bibinfo {volume} {12}},\ \bibinfo {pages} {4932} (\bibinfo {year}
  {2012})}\BibitemShut {NoStop}%
\bibitem [{\citenamefont {Genevet}\ \emph {et~al.}(2012)\citenamefont
  {Genevet}, \citenamefont {Yu}, \citenamefont {Aieta}, \citenamefont {Lin},
  \citenamefont {Kats}, \citenamefont {Blanchard}, \citenamefont {Scully},
  \citenamefont {Gaburro},\ and\ \citenamefont {Capasso}}]{Genevet2012}%
  \BibitemOpen
  \bibfield  {author} {\bibinfo {author} {\bibfnamefont {P.}~\bibnamefont
  {Genevet}}, \bibinfo {author} {\bibfnamefont {N.}~\bibnamefont {Yu}},
  \bibinfo {author} {\bibfnamefont {F.}~\bibnamefont {Aieta}}, \bibinfo
  {author} {\bibfnamefont {J.}~\bibnamefont {Lin}}, \bibinfo {author}
  {\bibfnamefont {M.~A.}\ \bibnamefont {Kats}}, \bibinfo {author}
  {\bibfnamefont {R.}~\bibnamefont {Blanchard}}, \bibinfo {author}
  {\bibfnamefont {M.~O.}\ \bibnamefont {Scully}}, \bibinfo {author}
  {\bibfnamefont {Z.}~\bibnamefont {Gaburro}}, \ and\ \bibinfo {author}
  {\bibfnamefont {F.}~\bibnamefont {Capasso}},\ }\href {\doibase
  10.1063/1.3673334} {\bibfield  {journal} {\bibinfo  {journal} {Applied
  Physics Letters}\ }\textbf {\bibinfo {volume} {100}},\ \bibinfo {pages}
  {013101} (\bibinfo {year} {2012})}\BibitemShut {NoStop}%
\bibitem [{\citenamefont {Lin}\ \emph {et~al.}(2014)\citenamefont {Lin},
  \citenamefont {Fan}, \citenamefont {Hasman},\ and\ \citenamefont
  {Brongersma}}]{Lin2014a}%
  \BibitemOpen
  \bibfield  {author} {\bibinfo {author} {\bibfnamefont {D.}~\bibnamefont
  {Lin}}, \bibinfo {author} {\bibfnamefont {P.}~\bibnamefont {Fan}}, \bibinfo
  {author} {\bibfnamefont {E.}~\bibnamefont {Hasman}}, \ and\ \bibinfo {author}
  {\bibfnamefont {M.~L.}\ \bibnamefont {Brongersma}},\ }\href {\doibase
  10.1126/science.1253213} {\bibfield  {journal} {\bibinfo  {journal}
  {Science}\ }\textbf {\bibinfo {volume} {345}},\ \bibinfo {pages} {298}
  (\bibinfo {year} {2014})}\BibitemShut {NoStop}%
\bibitem [{\citenamefont {Vo}\ \emph {et~al.}(2014)\citenamefont {Vo},
  \citenamefont {Fattal}, \citenamefont {Sorin}, \citenamefont {Peng},
  \citenamefont {Tran}, \citenamefont {Beausoleil},\ and\ \citenamefont
  {Fiorentino}}]{Vo2014}%
  \BibitemOpen
  \bibfield  {author} {\bibinfo {author} {\bibfnamefont {S.}~\bibnamefont
  {Vo}}, \bibinfo {author} {\bibfnamefont {D.}~\bibnamefont {Fattal}}, \bibinfo
  {author} {\bibfnamefont {W.~V.}\ \bibnamefont {Sorin}}, \bibinfo {author}
  {\bibfnamefont {Z.}~\bibnamefont {Peng}}, \bibinfo {author} {\bibfnamefont
  {T.}~\bibnamefont {Tran}}, \bibinfo {author} {\bibfnamefont {R.~G.}\
  \bibnamefont {Beausoleil}}, \ and\ \bibinfo {author} {\bibfnamefont
  {M.}~\bibnamefont {Fiorentino}},\ }\href {\doibase 10.1109/LPT.2014.2325947}
  {\bibfield  {journal} {\bibinfo  {journal} {IEEE Photonics Technology
  Letters}\ }\textbf {\bibinfo {volume} {26}},\ \bibinfo {pages} {1} (\bibinfo
  {year} {2014})}\BibitemShut {NoStop}%
\bibitem [{\citenamefont {Stork}\ \emph {et~al.}(1991)\citenamefont {Stork},
  \citenamefont {Streibl}, \citenamefont {Haidner},\ and\ \citenamefont
  {Kipfer}}]{Stork1991}%
  \BibitemOpen
  \bibfield  {author} {\bibinfo {author} {\bibfnamefont {W.}~\bibnamefont
  {Stork}}, \bibinfo {author} {\bibfnamefont {N.}~\bibnamefont {Streibl}},
  \bibinfo {author} {\bibfnamefont {H.}~\bibnamefont {Haidner}}, \ and\
  \bibinfo {author} {\bibfnamefont {P.}~\bibnamefont {Kipfer}},\ }\href
  {http://www.ncbi.nlm.nih.gov/pubmed/19784181} {\bibfield  {journal} {\bibinfo
   {journal} {Optics letters}\ }\textbf {\bibinfo {volume} {16}},\ \bibinfo
  {pages} {1921} (\bibinfo {year} {1991})}\BibitemShut {NoStop}%
\bibitem [{\citenamefont {Chen}\ and\ \citenamefont
  {Craighead}(1995)}]{Chen1995}%
  \BibitemOpen
  \bibfield  {author} {\bibinfo {author} {\bibfnamefont {F.~T.}\ \bibnamefont
  {Chen}}\ and\ \bibinfo {author} {\bibfnamefont {H.~G.}\ \bibnamefont
  {Craighead}},\ }\href {\doibase 10.1364/OL.20.000121} {\bibfield  {journal}
  {\bibinfo  {journal} {Optics Letters}\ }\textbf {\bibinfo {volume} {20}},\
  \bibinfo {pages} {121} (\bibinfo {year} {1995})}\BibitemShut {NoStop}%
\bibitem [{\citenamefont {Warren}\ \emph {et~al.}(1995)\citenamefont {Warren},
  \citenamefont {Smith}, \citenamefont {Vawter},\ and\ \citenamefont
  {Wendt}}]{Warren1995}%
  \BibitemOpen
  \bibfield  {author} {\bibinfo {author} {\bibfnamefont {M.~E.}\ \bibnamefont
  {Warren}}, \bibinfo {author} {\bibfnamefont {R.~E.}\ \bibnamefont {Smith}},
  \bibinfo {author} {\bibfnamefont {G.~A.}\ \bibnamefont {Vawter}}, \ and\
  \bibinfo {author} {\bibfnamefont {J.~R.}\ \bibnamefont {Wendt}},\ }\href
  {\doibase 10.1364/OL.20.001441} {\bibfield  {journal} {\bibinfo  {journal}
  {Optics Letters}\ }\textbf {\bibinfo {volume} {20}},\ \bibinfo {pages} {1441}
  (\bibinfo {year} {1995})}\BibitemShut {NoStop}%
\bibitem [{\citenamefont {Chen}\ and\ \citenamefont
  {Craighead}(1996)}]{Chen1996}%
  \BibitemOpen
  \bibfield  {author} {\bibinfo {author} {\bibfnamefont {F.~T.}\ \bibnamefont
  {Chen}}\ and\ \bibinfo {author} {\bibfnamefont {H.~G.}\ \bibnamefont
  {Craighead}},\ }\href {\doibase 10.1364/OL.21.000177} {\bibfield  {journal}
  {\bibinfo  {journal} {Optics Letters}\ }\textbf {\bibinfo {volume} {21}},\
  \bibinfo {pages} {177} (\bibinfo {year} {1996})}\BibitemShut {NoStop}%
\bibitem [{\citenamefont {Lalanne}\ \emph {et~al.}(1998)\citenamefont
  {Lalanne}, \citenamefont {Astilean}, \citenamefont {Chavel}, \citenamefont
  {Cambril},\ and\ \citenamefont {Launois}}]{Lalanne1998}%
  \BibitemOpen
  \bibfield  {author} {\bibinfo {author} {\bibfnamefont {P.}~\bibnamefont
  {Lalanne}}, \bibinfo {author} {\bibfnamefont {S.}~\bibnamefont {Astilean}},
  \bibinfo {author} {\bibfnamefont {P.}~\bibnamefont {Chavel}}, \bibinfo
  {author} {\bibfnamefont {E.}~\bibnamefont {Cambril}}, \ and\ \bibinfo
  {author} {\bibfnamefont {H.}~\bibnamefont {Launois}},\ }\href {\doibase
  10.1364/OL.23.001081} {\bibfield  {journal} {\bibinfo  {journal} {Optics
  Letters}\ }\textbf {\bibinfo {volume} {23}},\ \bibinfo {pages} {1081}
  (\bibinfo {year} {1998})}\BibitemShut {NoStop}%
\bibitem [{\citenamefont {Huang}\ \emph {et~al.}(2008)\citenamefont {Huang},
  \citenamefont {Kao}, \citenamefont {Fedotov}, \citenamefont {Chen},\ and\
  \citenamefont {Zheludev}}]{Huang2008}%
  \BibitemOpen
  \bibfield  {author} {\bibinfo {author} {\bibfnamefont {F.~M.}\ \bibnamefont
  {Huang}}, \bibinfo {author} {\bibfnamefont {T.~S.}\ \bibnamefont {Kao}},
  \bibinfo {author} {\bibfnamefont {V.~a.}\ \bibnamefont {Fedotov}}, \bibinfo
  {author} {\bibfnamefont {Y.}~\bibnamefont {Chen}}, \ and\ \bibinfo {author}
  {\bibfnamefont {N.~I.}\ \bibnamefont {Zheludev}},\ }\href {\doibase
  10.1021/nl801476v} {\bibfield  {journal} {\bibinfo  {journal} {Nano letters}\
  }\textbf {\bibinfo {volume} {8}},\ \bibinfo {pages} {2469} (\bibinfo {year}
  {2008})}\BibitemShut {NoStop}%
\bibitem [{\citenamefont {Ni}\ \emph {et~al.}(2013)\citenamefont {Ni},
  \citenamefont {Ishii}, \citenamefont {Kildishev},\ and\ \citenamefont
  {Shalaev}}]{Ni2013}%
  \BibitemOpen
  \bibfield  {author} {\bibinfo {author} {\bibfnamefont {X.}~\bibnamefont
  {Ni}}, \bibinfo {author} {\bibfnamefont {S.}~\bibnamefont {Ishii}}, \bibinfo
  {author} {\bibfnamefont {A.~V.}\ \bibnamefont {Kildishev}}, \ and\ \bibinfo
  {author} {\bibfnamefont {V.~M.}\ \bibnamefont {Shalaev}},\ }\href {\doibase
  10.1038/lsa.2013.28} {\bibfield  {journal} {\bibinfo  {journal} {Light:
  Science \& Applications}\ }\textbf {\bibinfo {volume} {2}},\ \bibinfo {pages}
  {e72} (\bibinfo {year} {2013})}\BibitemShut {NoStop}%
\bibitem [{\citenamefont {Karimi}\ \emph {et~al.}(2014)\citenamefont {Karimi},
  \citenamefont {Schulz}, \citenamefont {{De Leon}}, \citenamefont {Qassim},
  \citenamefont {Upham},\ and\ \citenamefont {Boyd}}]{Karimi2014}%
  \BibitemOpen
  \bibfield  {author} {\bibinfo {author} {\bibfnamefont {E.}~\bibnamefont
  {Karimi}}, \bibinfo {author} {\bibfnamefont {S.~A.}\ \bibnamefont {Schulz}},
  \bibinfo {author} {\bibfnamefont {I.}~\bibnamefont {{De Leon}}}, \bibinfo
  {author} {\bibfnamefont {H.}~\bibnamefont {Qassim}}, \bibinfo {author}
  {\bibfnamefont {J.}~\bibnamefont {Upham}}, \ and\ \bibinfo {author}
  {\bibfnamefont {R.~W.}\ \bibnamefont {Boyd}},\ }\href {\doibase
  10.1038/lsa.2014.48} {\bibfield  {journal} {\bibinfo  {journal} {Light:
  Science \& Applications}\ }\textbf {\bibinfo {volume} {3}},\ \bibinfo {pages}
  {e167} (\bibinfo {year} {2014})}\BibitemShut {NoStop}%
\bibitem [{\citenamefont {Monticone}\ \emph {et~al.}(2013)\citenamefont
  {Monticone}, \citenamefont {Estakhri},\ and\ \citenamefont
  {Al\`{u}}}]{Monticone2013}%
  \BibitemOpen
  \bibfield  {author} {\bibinfo {author} {\bibfnamefont {F.}~\bibnamefont
  {Monticone}}, \bibinfo {author} {\bibfnamefont {N.~M.}\ \bibnamefont
  {Estakhri}}, \ and\ \bibinfo {author} {\bibfnamefont {A.}~\bibnamefont
  {Al\`{u}}},\ }\href {\doibase 10.1103/PhysRevLett.110.203903} {\bibfield
  {journal} {\bibinfo  {journal} {Physical Review Letters}\ }\textbf {\bibinfo
  {volume} {110}},\ \bibinfo {pages} {203903} (\bibinfo {year}
  {2013})}\BibitemShut {NoStop}%
\bibitem [{\citenamefont {Chen}\ \emph {et~al.}(2006)\citenamefont {Chen},
  \citenamefont {Huang}, \citenamefont {Mateus}, \citenamefont
  {Chang-Hasnain},\ and\ \citenamefont {Suzuki}}]{Chen2006}%
  \BibitemOpen
  \bibfield  {author} {\bibinfo {author} {\bibfnamefont {L.}~\bibnamefont
  {Chen}}, \bibinfo {author} {\bibfnamefont {M.~C.~Y.}\ \bibnamefont {Huang}},
  \bibinfo {author} {\bibfnamefont {C.~F.~R.}\ \bibnamefont {Mateus}}, \bibinfo
  {author} {\bibfnamefont {C.~J.}\ \bibnamefont {Chang-Hasnain}}, \ and\
  \bibinfo {author} {\bibfnamefont {Y.}~\bibnamefont {Suzuki}},\ }\href
  {\doibase 10.1063/1.2164920} {\bibfield  {journal} {\bibinfo  {journal}
  {Applied Physics Letters}\ }\textbf {\bibinfo {volume} {88}},\ \bibinfo
  {pages} {031102} (\bibinfo {year} {2006})}\BibitemShut {NoStop}%
\bibitem [{\citenamefont {Kemiktarak}\ \emph {et~al.}(2012)\citenamefont
  {Kemiktarak}, \citenamefont {Metcalfe}, \citenamefont {Durand},\ and\
  \citenamefont {Lawall}}]{Kemiktarak2012}%
  \BibitemOpen
  \bibfield  {author} {\bibinfo {author} {\bibfnamefont {U.}~\bibnamefont
  {Kemiktarak}}, \bibinfo {author} {\bibfnamefont {M.}~\bibnamefont
  {Metcalfe}}, \bibinfo {author} {\bibfnamefont {M.}~\bibnamefont {Durand}}, \
  and\ \bibinfo {author} {\bibfnamefont {J.}~\bibnamefont {Lawall}},\ }\href
  {\doibase 10.1063/1.3684248} {\bibfield  {journal} {\bibinfo  {journal}
  {Applied Physics Letters}\ }\textbf {\bibinfo {volume} {100}},\ \bibinfo
  {pages} {061124} (\bibinfo {year} {2012})}\BibitemShut {NoStop}%
\bibitem [{\citenamefont {Wu}\ \emph {et~al.}(2012)\citenamefont {Wu},
  \citenamefont {Syu}, \citenamefont {Wu}, \citenamefont {Chen}, \citenamefont
  {Lu}, \citenamefont {Wang}, \citenamefont {Chiang},\ and\ \citenamefont
  {Tsai}}]{Wu2012a}%
  \BibitemOpen
  \bibfield  {author} {\bibinfo {author} {\bibfnamefont {T.~T.}\ \bibnamefont
  {Wu}}, \bibinfo {author} {\bibfnamefont {Y.~C.}\ \bibnamefont {Syu}},
  \bibinfo {author} {\bibfnamefont {S.~H.}\ \bibnamefont {Wu}}, \bibinfo
  {author} {\bibfnamefont {W.~T.}\ \bibnamefont {Chen}}, \bibinfo {author}
  {\bibfnamefont {T.~C.}\ \bibnamefont {Lu}}, \bibinfo {author} {\bibfnamefont
  {S.~C.}\ \bibnamefont {Wang}}, \bibinfo {author} {\bibfnamefont {H.~P.}\
  \bibnamefont {Chiang}}, \ and\ \bibinfo {author} {\bibfnamefont {D.~P.}\
  \bibnamefont {Tsai}},\ }\href {\doibase 10.1364/OE.20.020551} {\bibfield
  {journal} {\bibinfo  {journal} {Optics express}\ }\textbf {\bibinfo {volume}
  {20}},\ \bibinfo {pages} {20551} (\bibinfo {year} {2012})}\BibitemShut
  {NoStop}%
\bibitem [{\citenamefont {Mateus}\ \emph {et~al.}(2004)\citenamefont {Mateus},
  \citenamefont {Huang}, \citenamefont {Deng}, \citenamefont {Neureuther},\
  and\ \citenamefont {Chang-Hasnain}}]{Mateus2004}%
  \BibitemOpen
  \bibfield  {author} {\bibinfo {author} {\bibfnamefont {C.}~\bibnamefont
  {Mateus}}, \bibinfo {author} {\bibfnamefont {M.}~\bibnamefont {Huang}},
  \bibinfo {author} {\bibfnamefont {Y.}~\bibnamefont {Deng}}, \bibinfo {author}
  {\bibfnamefont {A.}~\bibnamefont {Neureuther}}, \ and\ \bibinfo {author}
  {\bibfnamefont {C.}~\bibnamefont {Chang-Hasnain}},\ }\href {\doibase
  10.1109/LPT.2003.821258} {\bibfield  {journal} {\bibinfo  {journal} {IEEE
  Photonics Technology Letters}\ }\textbf {\bibinfo {volume} {16}},\ \bibinfo
  {pages} {518} (\bibinfo {year} {2004})}\BibitemShut {NoStop}%
\bibitem [{\citenamefont {Lu}\ \emph {et~al.}(2010)\citenamefont {Lu},
  \citenamefont {Sedgwick}, \citenamefont {Karagodsky}, \citenamefont {Chase},\
  and\ \citenamefont {Chang-Hasnain}}]{Lu2010}%
  \BibitemOpen
  \bibfield  {author} {\bibinfo {author} {\bibfnamefont {F.}~\bibnamefont
  {Lu}}, \bibinfo {author} {\bibfnamefont {F.~G.}\ \bibnamefont {Sedgwick}},
  \bibinfo {author} {\bibfnamefont {V.}~\bibnamefont {Karagodsky}}, \bibinfo
  {author} {\bibfnamefont {C.}~\bibnamefont {Chase}}, \ and\ \bibinfo {author}
  {\bibfnamefont {C.~J.}\ \bibnamefont {Chang-Hasnain}},\ }\href {\doibase
  10.1364/OE.18.012606} {\bibfield  {journal} {\bibinfo  {journal} {Optics
  express}\ }\textbf {\bibinfo {volume} {18}},\ \bibinfo {pages} {12606}
  (\bibinfo {year} {2010})}\BibitemShut {NoStop}%
\bibitem [{\citenamefont {Astilean}\ \emph {et~al.}(1998)\citenamefont
  {Astilean}, \citenamefont {Lalanne}, \citenamefont {Chavel}, \citenamefont
  {Cambril},\ and\ \citenamefont {Launois}}]{Astilean1998}%
  \BibitemOpen
  \bibfield  {author} {\bibinfo {author} {\bibfnamefont {S.}~\bibnamefont
  {Astilean}}, \bibinfo {author} {\bibfnamefont {P.}~\bibnamefont {Lalanne}},
  \bibinfo {author} {\bibfnamefont {P.}~\bibnamefont {Chavel}}, \bibinfo
  {author} {\bibfnamefont {E.}~\bibnamefont {Cambril}}, \ and\ \bibinfo
  {author} {\bibfnamefont {H.}~\bibnamefont {Launois}},\ }\href {\doibase
  10.1364/OL.23.000552} {\bibfield  {journal} {\bibinfo  {journal} {Optics
  Letters}\ }\textbf {\bibinfo {volume} {23}},\ \bibinfo {pages} {552}
  (\bibinfo {year} {1998})}\BibitemShut {NoStop}%
\bibitem [{\citenamefont {Fattal}\ \emph {et~al.}(2010)\citenamefont {Fattal},
  \citenamefont {Li}, \citenamefont {Peng}, \citenamefont {Fiorentino},\ and\
  \citenamefont {Beausoleil}}]{Fattal2010}%
  \BibitemOpen
  \bibfield  {author} {\bibinfo {author} {\bibfnamefont {D.}~\bibnamefont
  {Fattal}}, \bibinfo {author} {\bibfnamefont {J.}~\bibnamefont {Li}}, \bibinfo
  {author} {\bibfnamefont {Z.}~\bibnamefont {Peng}}, \bibinfo {author}
  {\bibfnamefont {M.}~\bibnamefont {Fiorentino}}, \ and\ \bibinfo {author}
  {\bibfnamefont {R.~G.}\ \bibnamefont {Beausoleil}},\ }\href {\doibase
  10.1038/nphoton.2010.116} {\bibfield  {journal} {\bibinfo  {journal} {Nature
  Photonics}\ }\textbf {\bibinfo {volume} {4}},\ \bibinfo {pages} {466}
  (\bibinfo {year} {2010})}\BibitemShut {NoStop}%
\bibitem [{\citenamefont {Klemm}\ \emph {et~al.}(2013)\citenamefont {Klemm},
  \citenamefont {Stellinga}, \citenamefont {Martins}, \citenamefont {Lewis},
  \citenamefont {Huyet}, \citenamefont {Faolain}, \citenamefont {Krauss},\ and\
  \citenamefont {O'Faolain}}]{Klemm2013}%
  \BibitemOpen
  \bibfield  {author} {\bibinfo {author} {\bibfnamefont {A.~B.}\ \bibnamefont
  {Klemm}}, \bibinfo {author} {\bibfnamefont {D.}~\bibnamefont {Stellinga}},
  \bibinfo {author} {\bibfnamefont {E.~R.}\ \bibnamefont {Martins}}, \bibinfo
  {author} {\bibfnamefont {L.}~\bibnamefont {Lewis}}, \bibinfo {author}
  {\bibfnamefont {G.}~\bibnamefont {Huyet}}, \bibinfo {author} {\bibfnamefont
  {L.~O.}\ \bibnamefont {Faolain}}, \bibinfo {author} {\bibfnamefont {T.~F.}\
  \bibnamefont {Krauss}}, \ and\ \bibinfo {author} {\bibfnamefont
  {L.}~\bibnamefont {O'Faolain}},\ }\href {\doibase 10.1364/OL.38.003410}
  {\bibfield  {journal} {\bibinfo  {journal} {Optics letters}\ }\textbf
  {\bibinfo {volume} {38}},\ \bibinfo {pages} {3410} (\bibinfo {year}
  {2013})}\BibitemShut {NoStop}%
\bibitem [{\citenamefont {Arbabi}\ \emph
  {et~al.}(2014{\natexlab{a}})\citenamefont {Arbabi}, \citenamefont {Bagheri},
  \citenamefont {Ball}, \citenamefont {Horie}, \citenamefont {Fattal},\ and\
  \citenamefont {Faraon}}]{Arbabi2014}%
  \BibitemOpen
  \bibfield  {author} {\bibinfo {author} {\bibfnamefont {A.}~\bibnamefont
  {Arbabi}}, \bibinfo {author} {\bibfnamefont {M.}~\bibnamefont {Bagheri}},
  \bibinfo {author} {\bibfnamefont {A.~J.}\ \bibnamefont {Ball}}, \bibinfo
  {author} {\bibfnamefont {Y.}~\bibnamefont {Horie}}, \bibinfo {author}
  {\bibfnamefont {D.}~\bibnamefont {Fattal}}, \ and\ \bibinfo {author}
  {\bibfnamefont {A.}~\bibnamefont {Faraon}},\ }in\ \href
  {http://www.opticsinfobase.org/abstract.cfm?URI=CLEO\_SI-2014-STu3M.4} {\emph
  {\bibinfo {booktitle} {CLEO: 2014}}}\ (\bibinfo  {publisher} {Optical Society
  of America},\ \bibinfo {address} {San Jose, California},\ \bibinfo {year}
  {2014})\ p.\ \bibinfo {pages} {STu3M.4}\BibitemShut {NoStop}%
\bibitem [{\citenamefont {Shiono}\ \emph {et~al.}(1989)\citenamefont {Shiono},
  \citenamefont {Kitagawa}, \citenamefont {Setsune},\ and\ \citenamefont
  {Mitsuyu}}]{Shiono1989}%
  \BibitemOpen
  \bibfield  {author} {\bibinfo {author} {\bibfnamefont {T.}~\bibnamefont
  {Shiono}}, \bibinfo {author} {\bibfnamefont {M.}~\bibnamefont {Kitagawa}},
  \bibinfo {author} {\bibfnamefont {K.}~\bibnamefont {Setsune}}, \ and\
  \bibinfo {author} {\bibfnamefont {T.}~\bibnamefont {Mitsuyu}},\ }\href
  {\doibase 10.1364/AO.28.003434} {\bibfield  {journal} {\bibinfo  {journal}
  {Applied optics}\ }\textbf {\bibinfo {volume} {28}},\ \bibinfo {pages} {3434}
  (\bibinfo {year} {1989})}\BibitemShut {NoStop}%
\bibitem [{\citenamefont {Haruna}\ \emph {et~al.}(1990)\citenamefont {Haruna},
  \citenamefont {Takahashi}, \citenamefont {Wakahayashi},\ and\ \citenamefont
  {Nishihara}}]{Haruna1990}%
  \BibitemOpen
  \bibfield  {author} {\bibinfo {author} {\bibfnamefont {M.}~\bibnamefont
  {Haruna}}, \bibinfo {author} {\bibfnamefont {M.}~\bibnamefont {Takahashi}},
  \bibinfo {author} {\bibfnamefont {K.}~\bibnamefont {Wakahayashi}}, \ and\
  \bibinfo {author} {\bibfnamefont {H.}~\bibnamefont {Nishihara}},\ }\href
  {\doibase 10.1364/AO.29.005120} {\bibfield  {journal} {\bibinfo  {journal}
  {Applied optics}\ }\textbf {\bibinfo {volume} {29}},\ \bibinfo {pages} {5120}
  (\bibinfo {year} {1990})}\BibitemShut {NoStop}%
\bibitem [{\citenamefont {Oskooi}\ \emph {et~al.}(2010)\citenamefont {Oskooi},
  \citenamefont {Roundy}, \citenamefont {Ibanescu}, \citenamefont {Bermel},
  \citenamefont {Joannopoulos},\ and\ \citenamefont {Johnson}}]{Oskooi2010b}%
  \BibitemOpen
  \bibfield  {author} {\bibinfo {author} {\bibfnamefont {A.~F.}\ \bibnamefont
  {Oskooi}}, \bibinfo {author} {\bibfnamefont {D.}~\bibnamefont {Roundy}},
  \bibinfo {author} {\bibfnamefont {M.}~\bibnamefont {Ibanescu}}, \bibinfo
  {author} {\bibfnamefont {P.}~\bibnamefont {Bermel}}, \bibinfo {author}
  {\bibfnamefont {J.}~\bibnamefont {Joannopoulos}}, \ and\ \bibinfo {author}
  {\bibfnamefont {S.~G.}\ \bibnamefont {Johnson}},\ }\href {\doibase
  10.1016/j.cpc.2009.11.008} {\bibfield  {journal} {\bibinfo  {journal}
  {Computer Physics Communications}\ }\textbf {\bibinfo {volume} {181}},\
  \bibinfo {pages} {687} (\bibinfo {year} {2010})}\BibitemShut {NoStop}%
\bibitem [{\citenamefont {Arbabi}\ \emph
  {et~al.}(2014{\natexlab{b}})\citenamefont {Arbabi}, \citenamefont {Horie},\
  and\ \citenamefont {Faraon}}]{Arbabi2014a}%
  \BibitemOpen
  \bibfield  {author} {\bibinfo {author} {\bibfnamefont {A.}~\bibnamefont
  {Arbabi}}, \bibinfo {author} {\bibfnamefont {Y.}~\bibnamefont {Horie}}, \
  and\ \bibinfo {author} {\bibfnamefont {A.}~\bibnamefont {Faraon}},\ }in\
  \href {http://www.opticsinfobase.org/abstract.cfm?URI=CLEO\_SI-2014-STu3M.5}
  {\emph {\bibinfo {booktitle} {CLEO: 2014}}}\ (\bibinfo  {publisher} {Optical
  Society of America},\ \bibinfo {address} {San Jose, California},\ \bibinfo
  {year} {2014})\ p.\ \bibinfo {pages} {STu3M.5}\BibitemShut {NoStop}%
\end{thebibliography}

\begin{thebibliography}{3}%
\makeatletter
\providecommand \@ifxundefined [1]{%
 \@ifx{#1\undefined}
}%
\providecommand \@ifnum [1]{%
 \ifnum #1\expandafter \@firstoftwo
 \else \expandafter \@secondoftwo
 \fi
}%
\providecommand \@ifx [1]{%
 \ifx #1\expandafter \@firstoftwo
 \else \expandafter \@secondoftwo
 \fi
}%
\providecommand \natexlab [1]{#1}%
\providecommand \enquote  [1]{``#1''}%
\providecommand \bibnamefont  [1]{#1}%
\providecommand \bibfnamefont [1]{#1}%
\providecommand \citenamefont [1]{#1}%
\providecommand \href@noop [0]{\@secondoftwo}%
\providecommand \href [0]{\begingroup \@sanitize@url \@href}%
\providecommand \@href[1]{\@@startlink{#1}\@@href}%
\providecommand \@@href[1]{\endgroup#1\@@endlink}%
\providecommand \@sanitize@url [0]{\catcode `\\12\catcode `\$12\catcode
  `\&12\catcode `\#12\catcode `\^12\catcode `\_12\catcode `\%12\relax}%
\providecommand \@@startlink[1]{}%
\providecommand \@@endlink[0]{}%
\providecommand \url  [0]{\begingroup\@sanitize@url \@url }%
\providecommand \@url [1]{\endgroup\@href {#1}{\urlprefix }}%
\providecommand \urlprefix  [0]{URL }%
\providecommand \Eprint [0]{\href }%
\providecommand \doibase [0]{http://dx.doi.org/}%
\providecommand \selectlanguage [0]{\@gobble}%
\providecommand \bibinfo  [0]{\@secondoftwo}%
\providecommand \bibfield  [0]{\@secondoftwo}%
\providecommand \translation [1]{[#1]}%
\providecommand \BibitemOpen [0]{}%
\providecommand \bibitemStop [0]{}%
\providecommand \bibitemNoStop [0]{.\EOS\space}%
\providecommand \EOS [0]{\spacefactor3000\relax}%
\providecommand \BibitemShut  [1]{\csname bibitem#1\endcsname}%
\let\auto@bib@innerbib\@empty
\bibitem [{\citenamefont {Harrington}(2001)}]{Harrington2001}%
  \BibitemOpen
  \bibfield  {author} {\bibinfo {author} {\bibfnamefont {R.~F.}\ \bibnamefont
  {Harrington}},\ }\href
  {http://www.wiley.com/WileyCDA/WileyTitle/productCd-047120806X.html} {\emph
  {\bibinfo {title} {{Time Harmonic Electromagnetic Fields}}}}\ (\bibinfo
  {publisher} {Wiley-IEEE Press},\ \bibinfo {year} {2001})\BibitemShut
  {NoStop}%
\bibitem [{\citenamefont {Born}\ and\ \citenamefont {Wolf}(1999)}]{Born1999}%
  \BibitemOpen
  \bibfield  {author} {\bibinfo {author} {\bibfnamefont {M.}~\bibnamefont
  {Born}}\ and\ \bibinfo {author} {\bibfnamefont {E.}~\bibnamefont {Wolf}},\
  }\href@noop {} {\emph {\bibinfo {title} {{Principles of Optics}}}},\ \bibinfo
  {edition} {7th}\ ed.\ (\bibinfo  {publisher} {Cambridge University Press},\
  \bibinfo {address} {Cambridge},\ \bibinfo {year} {1999})\BibitemShut
  {NoStop}%
\bibitem [{\citenamefont {Wyrowski}(1992)}]{Wyrowski1992}%
  \BibitemOpen
  \bibfield  {author} {\bibinfo {author} {\bibfnamefont {F.}~\bibnamefont
  {Wyrowski}},\ }\href {\doibase 10.1016/0030-4018(92)90231-F} {\bibfield
  {journal} {\bibinfo  {journal} {Optics Communications}\ }\textbf {\bibinfo
  {volume} {92}},\ \bibinfo {pages} {119} (\bibinfo {year} {1992})}\BibitemShut
  {NoStop}%
\end{thebibliography}
\end{document}